\documentclass[aps,prd,amsmath,amssymb,twocolumn,preprintnumbers,nofootinbib]{revtex4-1}

\pdfoutput=1
\usepackage[utf8]{inputenc} 
\usepackage{graphicx}
\usepackage{multirow}
\usepackage{url}
\usepackage[bookmarks, pagebackref=false]{hyperref}
\usepackage[usenames,dvipsnames]{xcolor}
\definecolor{rossoCP3}{cmyk}{0,.88,.77,.40}
\definecolor{blaa}{rgb}{0.2,0.2,0.6}
\hypersetup{colorlinks, 
	bookmarksopen, 
	bookmarksnumbered,
	citecolor=blaa,
	linkcolor=rossoCP3,
	urlcolor=rossoCP3, 
}
\usepackage{bm}
\usepackage{bbm}

\usepackage{amsmath,amssymb,amsfonts,amsthm}
\usepackage{float}
\usepackage[Symbolsmallscale]{upgreek}
\usepackage{dsfont}
\usepackage[vcentermath]{youngtab}

\usepackage{placeins}
\usepackage{xspace}
\usepackage{cancel} 
\usepackage{slashed}
\usepackage{natbib}
\usepackage{booktabs,multirow}

\graphicspath{{./Figs/}}

\begin{document}


\title{\Large\color{rossoCP3}  Analytic Coupling Structure of Large \texorpdfstring{$N_f$}{N_f} (Super) QED and QCD}
\author{Nicola Andrea {\sc Dondi}$^{\color{rossoCP3}{\heartsuit}}$}
\author{Gerald V.~{\sc Dunne}$^{\color{rossoCP3}{\triangle}}$}
\author{Manuel {\sc Reichert}$^{\color{rossoCP3}{\heartsuit}}$}
\author{Francesco {\sc Sannino}$^{\color{rossoCP3}{\heartsuit\diamondsuit}}$}
\affiliation{\mbox{ $^{\color{rossoCP3}{\triangle}}$  Physics Department, University of Connecticut, Storrs CT 06269-3046, USA}\\\mbox{{ $^{\color{rossoCP3}{\heartsuit}}$ {CP}$^{ \bf 3}${-Origins}} \&  \rm{Danish IAS},  University of Southern Denmark, Campusvej 55, 5230 Odense M, Denmark}
 \\ $^{\color{rossoCP3}{\diamondsuit}}$ {SLAC, National Accelerator Laboratory, Stanford University, Stanford, CA 94025, USA}
}

\begin{abstract} 
We study the analytic properties of the 't Hooft coupling expansion of the beta-function at the leading nontrivial large-$N_f$ order for QED, QCD,  Super QED and Super QCD. For each theory, the 't Hooft coupling expansion  is convergent. We discover that an analysis of the expansion coefficients to roughly 30 orders is required to establish the radius of convergence accurately, and to characterize the (logarithmic) nature of the first singularity. We study summations of the beta-function expansion at order $1/N_f$ and identify the physical origin of the singularities in terms of iterated bubble diagrams. We find a common analytic structure across these theories, with important technical differences between supersymmetric and non-supersymmetric theories. We also discuss the expected structure at higher orders in the $1/N_f$ expansion, which will be in the future accessible with the methods presented in this work, meaning without the need for resumming the perturbative series. Understanding the structure of the large-$N_f$ expansion is an essential step towards determining the ultraviolet fate of asymptotically non-free gauge theories.  
\\
[.3cm]
{\footnotesize  \it Preprint: CP$^3$-Origins-2019-007 DNRF90
}

\end{abstract}

\maketitle

\section{Introduction}

The  discovery of four dimensional asymptotically safe quantum field theories  \cite{Litim:2014uca} has spurred recent phenomenological and theoretical interest. The original result made use of the Veneziano limit, in which one considers a large number of both colors and flavors. These theories feature perturbative safety and contain not only gauge and fermion degrees of freedom but also scalars. It is therefore theoretically and phenomenologically important to investigate the ultraviolet fate of non-asymptotically free gauge theories featuring a small number of colors but still a large number of flavors. In particular, one wishes to either exclude or demonstrate that a large number of flavors can lead to an asymptotically safe scenario in gauge-fermion theories.
 
This quest has revitalized the study of quantum field theories at a large number of flavors  $N_f$. 
The timeliness of our investigation is further corroborated by the fact that the large $N_f$ non-asymptotically free regime of gauge-fermion theories is being, for the first time, investigated via first principle lattice simulations where we expect the first results to appear soon \cite{LatticeNf}.

An intriguing property of this limit is that, at each order in the $1/N_f$ expansion, only a  finite number of underlying topologies contributes, where each gauge line is dressed with matter loops.   Correspondingly, at fixed order in $1/N_f$ the number of diagrams grows polynomially with the loop order, suggesting that a closed form resummed result with a finite radius of convergence may be achievable.

The first notable study is large-$N_f$ Quantum Electrodynamics (QED) \cite{PalanquesMestre:1983zy}, while large-$N_f$ Quantum Chromodynamics (QCD) was considered later in \cite{Gracey:1996he}. A historical summary of the techniques and earlier results can be found in \cite{Holdom:2010qs,Gracey:2018ame}. The generalization to a wide class of semi-simple gauge-Yukawa theories appeared only recently in \cite{Mann:2017wzh,Pelaggi:2017abg,Antipin:2018zdg,Alanne:2018ene,Alanne:2018csn,Kowalska:2017pkt}.  For gauge theories with different fermion matter representations the new phase diagram as a function of the number of flavours and colors was put forward in \cite{Antipin:2017ebo,Pica:2010xq}, and it was termed {\it Conformal Window 2.0}, extending and generalizing the original phase diagram of \cite{Sannino:2004qp,Dietrich:2006cm} to contain, besides an infrared conformal window, also an ultraviolet (safe) one. 

In the limit of a large number of matter fields, it is natural to introduce the 't Hooft coupling  
\begin{align}
K = \frac{g^2 N_f S_{2}(R)}{4\pi^2} \,,
\end{align}
with the gauge coupling $g$ and the Dynkin index $S_2(R)$, normalized to 1/2 for the fundamental representation. The generic beta-function has a formal expansion as an inverse series in $N_f$
\begin{align}
   \label{eq:beta}
  \beta(K)  &= \sum_{k=0}^\infty \frac{ \beta^{(k)}(K)}{N_f^k} \ ,
\end{align}
where each $\beta^{(k)}(K)$ has itself a perturbative expansion in the 't Hooft coupling $K$.
Similar expansions hold  for anomalous dimensions and other critical quantities. 

Asymptotic freedom is lost for theories at finite number of colors and large number of flavors, and therefore such theories can only be fundamental if they develop an interacting fixed point in the ultraviolet. This cannot occur in perturbation theory without Yukawa interactions \cite{Caswell:1974gg,Litim:2014uca}, but it may occur non-perturbatively above a critical number of flavors \cite{Antipin:2018zdg}. To see how this might work, let us schematically consider the leading  nontrivial order $1/N_f$ beta-function, which up to a normalization reads 
\begin{align}
\label{eq:cancel}
\frac{\beta(K)}{K^2} = 1 + \frac{1}{N_f} \frac{\beta^{(1)}(K)}{K^2}  + \mathcal{O}\!\left( \frac1{N_f^{2}}\right)\,.
\end{align}
At this order the function $\beta^{(1)} (K)/K^2$ must develop a singular behavior for the beta-function to develop a zero as $N_f\to\infty$. This indeed happens for QED and QCD, as summarized in \cite{Holdom:2010qs,Pica:2010xq,Antipin:2018zdg}. 
 
In this paper we investigate this phenomenon further and make a systematic study of the analytic structure of the 't Hooft  coupling expansion of the leading large-$N_f$  beta-function for QED, QCD, Super QED (SQED) and Super QCD (SQCD).  We discover that for each theory the 't Hooft coupling expansion is convergent, but a large number of expansion coefficients are needed in order to determine accurately the radius of convergence and to extract the logarithmic nature of the first singularity of the theory. Additionally, by a detailed investigation of the summation properties of the beta-function at leading order of $1/N_f$, we identify the physical origin of the singularities from the iterated self-energy diagrams. We find a universal analytic structure across the theories investigated here, while being able to resolve important physical differences between supersymmetric and non-supersymmetric theories. 

The paper is organized as follows: In \autoref{QED}, we investigate large-$N_f$ QED and introduce the relevant mathematical tests and tools that we use for the various theories. These include the asymptotic analysis of the expansion coefficients and Pad\'e approximants. We then identify  the physical origin of the poles. We extend this analysis to QCD, SQED, and SQCD in \autoref{sec:comparison}. There we also elucidate and highlight the crucial differences among the various theories. We present our conclusions in \autoref{sec:conclusions}. In App.~\ref{app:Darboux}, we briefly review  Darboux's theorem, relevant for the large-order behavior of the expansion coefficients, and in App.~\ref{BAppendix}, we describe on a technical level how we extracted the numerical coefficients of the beta-function. 
 
\section{Large \texorpdfstring{$N_f$}{N_f}  QED}
\label{QED}
\begin{figure}[t!]
\includegraphics[width=\linewidth]{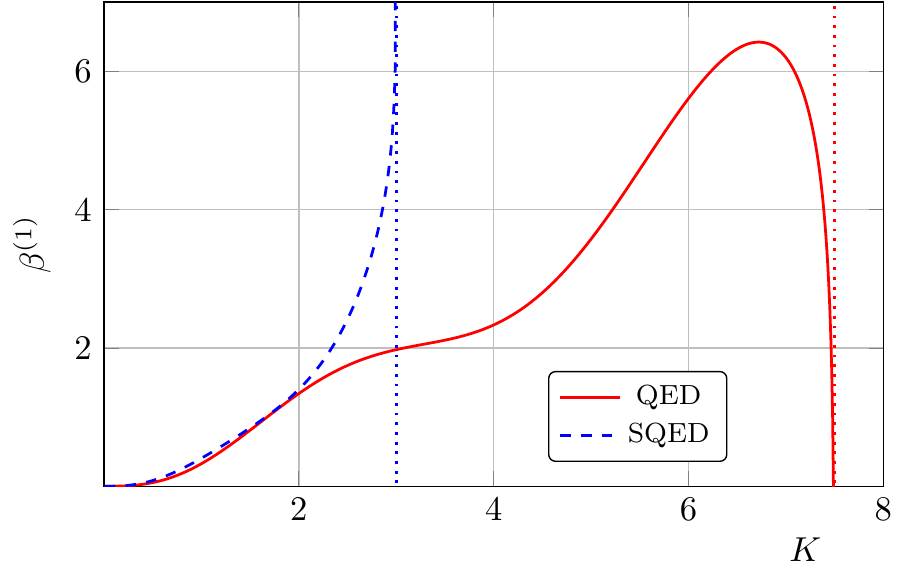}
\caption{Leading order $1/N_f$ beta-function for QED and SQED. The dotted vertical lines indicate the location of the first singular structure of each theory. Note the opposite sign of the behavior at the leading singularity: for SQED $\beta^{(1)}(K)\to +\infty$, while for QED $\beta^{(1)}(K)\to -\infty$.}
\label{fig:qed-sqed}
\end{figure}

QED is structurally the simplest gauge theory,  but it still has a rich perturbative and non-perturbative structure, which we probe here in the large $N_f$ limit. The QED beta-function has been computed in \cite{PalanquesMestre:1983zy} at the  leading non-trivial order in the $1/N_f$ expansion:
\begin{align}
   \label{eq:QED}
  \beta_\text{QED}(K)  &= \frac{2}{3}K^2 
  + \frac{K^2}{2N_f}  \int_0^K \!\!  \mathrm d x\, F_\text{QED} (x)+ \mathcal{O}\!\left(\frac{1}{N_f^2}\right)\, . 
\end{align}
Here the integrand function for QED is
\begin{align}
   \label{eq:QEDF}
  F_\text{QED}(x) &=-
   \frac{(x+3) (x - \frac92) ( x - \frac32) \sin\!\left(\frac{\pi  x}{3}\right) 
  \Gamma\!\left(\frac{5}{2}-\frac{x}{3} \right)}{27\cdot 2^{\frac{2 x}{3}-5} \pi ^{\frac32} (x-3) x \,\Gamma\!\left(3-\frac{x}{3}\right)}   \,. 
\end{align}
This resummed beta-function is shown in \autoref{fig:qed-sqed} compared with the supersymmetric version of the model. 

The first few terms in the $1/N_f$ term of \eqref{eq:QED} read
\begin{align}
   \label{eq:QED-leading}
  \beta^{(1)}_\text{QED}(K)  \equiv{}& 
     \frac{K^2}{2}\sum_{n=1}^\infty c_n^{(1)} K^n    \\
 ={}&  \frac{K^2}{2}\Big[  K - \frac{11}{36} K^2 - \frac{77}{972} K^3
   \notag\\
 & + \frac{107+144\zeta(3)}{7776} K^4 \notag\\
 &
 +\frac{1255+24 \pi ^4-2640 \zeta (3)}{291600} K^5+ \mathcal{O}(K^6)\Big] \, .
 \notag
\end{align}
In the coefficients $c_n^{(1)}$ we recognize characteristic $\pi$ powers and zeta values, familiar from algebraic properties of Feynman perturbation theory and harmonic polylogarithms \cite{Kreimer,Weinzierl:2003jx}.

Before discussing the analytic structure of the integral representation in \eqref{eq:QED}, consider the following pragmatic question: suppose, as is often the case, one were given only a finite number of terms of the expansion in \eqref{eq:QED-leading}, what could we learn about the physical nature of the expansion? There is a well-developed formalism to address such a question \cite{Fisher,Guttmann}. 
The first observation is that the expansion is convergent. This  can be confirmed by a variety of simple ratio tests: for example, the radius of convergence $c_*^{(1)}$ can be deduced from the limit $c_*^{(1)}=\lim_{n\to\infty}|c^{(1)}_n|^{-1/n}$, or from the limit $1/c_*^{(1)}=\lim_{n\to\infty}|c^{(1)}_n/c^{(1)}_{n-1}|$. However, more information about the physics of the expansion can be obtained by applying Darboux's theorem \cite{Fisher,Guttmann,Henrici}, which relates the rate of growth of the perturbative expansion coefficients to the behavior of the expansion about the leading and subleading singularities. Note that this is a stronger statement than simply saying that the {\it location} of the nearest singularity determines the radius of convergence.  The expansion coefficients also encode further information about the {\it nature} of the singularity. The general argument is summarized in App.~\ref{app:Darboux}.

\subsection{Asymptotic Analysis of Expansion Coefficients}
\label{sec:qed-asymptotics}
Our goal in this section is to deduce physical information from a finite number of expansion coefficients $c_n^{(1)}$ in 
\eqref{eq:QED-leading}. We studied these coefficients up to order $M=60$, and 
from 60 terms we obtain a great deal of asymptotic information. 
Using Richardson extrapolation \cite{Bender} with these 60 coefficients, we learn that as $n\to\infty$
\begin{align}
 \label{eq:large-order}
c_n^{(1)} \sim \frac{1}{(n+1)}&\left[ R_0\left(\frac{2}{15}\right)^{n+1} +R_1\left(\frac{2}{21}\right)^{n+1} \right. \notag \\
&\left. \quad+ R_2\left(\frac{2}{27}\right)^{n+1} +\dots\right] \, ,
\end{align} 
where $R_0=0.063044292$, $R_1=-0.013027009$, and $R_2=0.0033170626$. These numbers can be fit to $R_0=\frac{28}{45 \pi ^2}$, $R_1=-\frac{9}{70 \pi ^2}$, and $R_2=\frac{11}{336 \pi ^2}$, identifications that can be confirmed to higher precision using higher-order Richardson extrapolations. We explain the origin of these coefficients below in \eqref{eq:QED-pole-residue}.

\begin{figure}[t!]
\includegraphics[width=\linewidth]{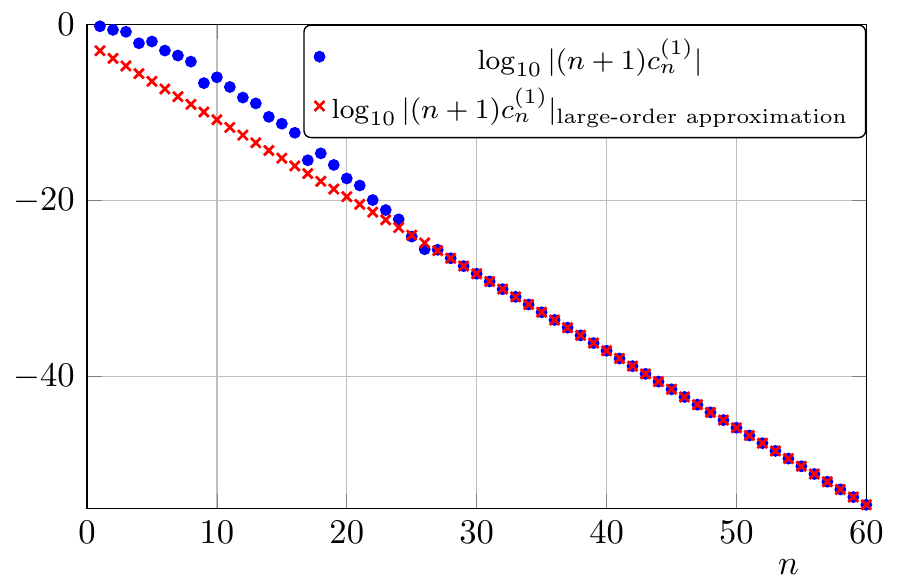}
\caption{Logarithm of  $|(n+1)\, c^{(1)}_n|$, for the expansion coefficients $c_n^{(1)}$ entering the leading $1/N_f$ QED beta-function $\beta^{(1)}(K)$ in \eqref{eq:QED-leading}, compared to the numerically extracted large-order behavior in \eqref{eq:large-order}. From $n\approx 30$ onwards, the coefficients agree with the expectation from the large-order behavior.
}
\label{fig:QED-coeffs}
\end{figure}

Thus, using Darboux's theorem (see App.~\ref{app:Darboux}), from these 60 perturbative expansion coefficients we learn that: (i) the radius of convergence is $15/2$; (ii) the leading singularity of $ \beta^{(1)}_\text{QED}(K)$ at $K_*=15/2$ is a logarithmic branch point, with coefficient $\frac12 R_0   K^2=\frac{14K^2}{45 \pi ^2}$; (iii) there are no higher-order corrections associated with this singularity; (iv) there are higher-order corrections associated with further singularities at $K=\frac{21}{2}$, and $K=\frac{27}{2}$.
Interestingly, we need approximately  $M=30$  terms of the expansion to be able to deduce precise  information about the leading singularity. With fewer than $M=30$ terms, even identifying the radius of convergence to be $15/2$ is noisy, see \autoref{fig:QED-coeffs}. To extract accurately the second and third singularities, and their coefficients, we require $M\approx 40$ and $M\approx 50$, respectively.

Thus, the leading behavior of the $1/N_f$ correction to the beta-function as $K$ approaches the radius of convergence is
\begin{align}
\label{eq:log}
\beta^{(1)}_\text{QED}(K)\sim &\frac{14 K^2}{45\pi^2 }\ln\left(\frac{15}{2}-K\right)+\dots \,\, , \quad 
K\to \frac{15}{2} \,.
\end{align}
This implies that in order to obtain a zero of the beta-function in the large $N_f$ limit, we must approach a non-perturbative fixed point at \cite{Litim:2014uca}
\begin{align}
K_*^\text{np}=\frac{15}{2}-\exp\!\left[-\frac{15\pi^2}{7}\, N_f\right]\,.
\label{eq:npk}
\end{align}
This physical information has been deduced from a finite number of terms in the perturbative expansion of $\beta^{(1)}(K)$. However,  since we have an all-orders integral representation \cite{PalanquesMestre:1983zy} in \eqref{eq:QED}, we can probe the analytic structure more precisely by studying the properties of the integrand function $F_\text{QED}(x)$ defined in \eqref{eq:QEDF}.
The singularities of the integrand are simple poles at $x_n=\frac{15}{2}+3n$, for $n\geq 0$, generated by  $\Gamma(\frac52-\frac{x}3)$. These are the only singularities, as can be seen from the decomposition
\begin{align}
 \Gamma\!\left( \frac{5}{2} - \frac{x}{3} \right) = 
 \Gamma\!\left( \frac{5}{2} - \frac{x}{3},1 \right) 
 + \sum_{n=0}^{\infty} \frac{(-1)^n}{n!} \frac{1}{\left(x - 3n - \frac{15}{2}\right)} \,,
\end{align}
where the incomplete gamma function $\Gamma\!\left( \frac{5}{2} - \frac{x}{3},1 \right)$ is regular. The potential poles at $x=3-3n$, with $n\geq 0$, coming from the denominator in \eqref{eq:QEDF} are in fact canceled by the $\sin\!\left(\frac{\pi x}{3}\right)$ factor in the numerator. 
Alternatively, one can re-write the integrand using the gamma function reflection formula as
\begin{align}
\label{eq:QED-alt}
F_\text{QED}(x) &=
 -  \left(\frac{ \sin\!\left(\frac{\pi  x}{3}\right) }{x(x-3)(x-6)}\right)^2 \left(\frac{(x - \frac92) ( x - \frac32)}{\cos\!\left(\frac{\pi  x}{3}\right)}\right)
\notag    \\
   &\quad \times  \frac{2^{5-\frac{2 x}{3}} \left(x-6\right) \left(x - \frac92\right) \left( x - \frac32\right)}{9\pi ^{3/2}}
  \frac{\Gamma\!\left(1+\frac{x}{3} \right)}{\Gamma\!\left(\frac{1}{2}+\frac{x}{3}\right)}   \, ,\end{align}
from which we see that the only singularities come from the ${\rm sec}\!\left(\frac{\pi  x}{3}\right)$  factor, with the poles at $x=\frac{3}{2}$ and $x=\frac{9}{2}$ excluded. 
Therefore, the positions and residues of the (simple) poles of the integrand are:
\begin{align}
 \label{eq:QED-pole-residue}
 x_n &= 3n + \frac{15}{2} \,,\qquad n=0,\, 1,\, 2,\, \dots
 \notag \\
 \text{R}_n &= \frac{2^{1-2n} (n+1) (n+2) (2 n+7)}{3 \pi ^{3/2} (2 n+3) (2 n+5)
 \,\Gamma\!\left(\frac{1}{2}-n\right) \Gamma\!\left(n+1\right)} \,.
\end{align}
These coincide precisely with the numerical values extracted from the asymptotic analysis in \eqref{eq:large-order}. Furthermore, the noisiness of the expansion coefficients at low order can be traced to the oscillatory nature of the $\sin\!\left(\frac{\pi  x}{3}\right) \tan\!\left(\frac{\pi  x}{3}\right)$ factor in \eqref{eq:QED-alt}.

Recall that the poles in \eqref{eq:QED-pole-residue} are simple poles of the {\it integrand} of the beta-function in \eqref{eq:QED}. After integration over $x$, these poles translate into {\it logarithmic} branch points of the beta-function, which were found above [see, e.g., \eqref{eq:log}] by a numerical Darboux analysis of the coefficients of the perturbative expansion of the beta-function to finite order.

\subsection{Pad\'e Approximations}
\label{sec:pade}
Pad\'e approximation is a commonly used method for studying perturbative expansions in physical systems \cite{Baker,Bender}. 
Given the integrand $F_\text{QED}(x)$ in \eqref{eq:QEDF} expressed in terms of gamma functions, there is a {\it unique} analytic continuation beyond its radius of convergence. However, if we only had a finite number of terms of the expansion, not its full analytic form, we could still probe  beyond the radius of convergence using Pad\'e approximation. 

Pad\'e approximants construct analytic continuations of truncated Taylor series (i.e., polynomial) approximations to functions, expressing the given polynomial as a ratio of two polynomials of lower order, with coefficients determined purely algorithmically. Pad\'e approximants thus convert a polynomial to a rational function, which can also be expressed as a partial fraction expansion, whose residues and poles are determined by the coefficients of the original truncated Taylor series. This means that Pad\'e approximants tend to be quite good at representing function with poles, while they are less good at representing functions with branch cuts \cite{Baker,Bender}.

The conversion of a truncated Taylor series to a Pad\'e approximant
\begin{align}
\label{eq:pade}
 F_\text{QED} (x) \approx   \sum_{n=0}^{M} f_n \,x^n \quad \longrightarrow \quad
 \mathcal P^{[R, S]}(x)=\frac{P_R(x)}{Q_S(x)} \,,
\end{align}
is  algorithmic, leading to a ratio of two polynomials $P_R(x)$ and $Q_S(x)$, of order $R$ and $S$ respectively, where $R+S=M$. It is in fact a built-in function in symbolic mathematics languages such as Maple or Mathematica.

\begin{figure}[t!]
\includegraphics[width=\linewidth]{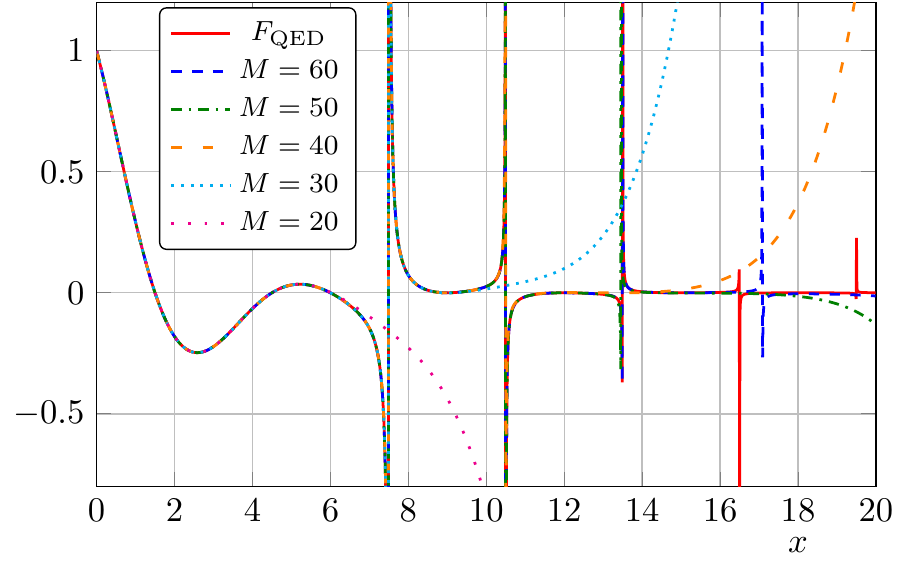}
\caption{Integrand of the QED beta-function $F_\text{QED}(x)$ (solid red curve), see \eqref{eq:QED},
 compared to the diagonal  Pad\'e approximant $\mathcal P^{[M/2,M/2]}$  for $M$ terms in the perturbative expansion in \eqref{eq:pade}. Progressively more poles are resolved as the order of the Pad\'e approximant is increased.
}
\label{fig:qed-full-vs-pade}
\end{figure}

We took up to 60 terms of the expansion about $x=0$ of the integrand $F_\text{QED}(x)$, and converted it to a diagonal Pad\'e approximant $\mathcal P^{[M/2,M/2]}(x)$, for various values of $M$.
In \autoref{fig:qed-full-vs-pade} we display the function $F_\text{QED}(x)$ together
with the diagonal Pad\'e approximants starting from $M=20,\, 30,\, 40,\, 50,\, 60$ coefficients.
With $M=20$ coefficients we do not even "see" the first pole.
With $M=30$ coefficients we accurately probe the first pole, but not the second pole.
For the second pole we need approximately $M=40$ coefficients,
while with $M=50$ coefficients we accurately resolve the third pole.
These numbers are consistent with the number of coefficients required in the ratio test and asymptotic analysis of the beta-function coefficients in \autoref{sec:qed-asymptotics}, to resolve the logarithmic singularities of $\beta^{(1)}(K)$. 

In fact, a full Pad\'e analysis constructs the ``Pad\'e table'', of all Pad\'e approximants $\mathcal P^{[R, S]}(x)$, with $R+S=M$. It turns out that certain off-diagonal approximants are even better at representing the integrand function $F_\text{QED}(x)$. This can be understood from the analytic representation of the integrand in \eqref{eq:QED-alt}.  Given that trigonometric and gamma  functions have well-known product formula representations, we see that the $\Gamma\!\left(1+\frac{x}{3} \right)/\Gamma\!\left(\frac{1}{2}+\frac{x}{3}\right)$ factor in \eqref{eq:QED-alt} is naturally represented as a near-diagonal Pad\'e approximant, but because of the $\sin^2\!\left(\frac{\pi  x}{3}\right)/\cos\!\left(\frac{\pi  x}{3}\right)$ factor, there is effectively one extra trigonometric factor in the numerator. Thus a Pad\'e representation whose numerator is a higher-order polynomial than the denominator polynomial will represent the analytic structure of the actual function $F_\text{QED}(x)$ more accurately. We have confirmed that this is the case, starting from the $60$ expansion coefficients, but we note that the simple diagonal Pad\'e representations shown in \autoref{fig:qed-full-vs-pade} are already remarkably precise.

\subsection{Physical Origin of the Poles}
\label{sec:origin}
We have seen that the finite radius of convergence, $K_*=\frac{15}{2}$, of the expansion of the $1/N_f$ beta-function $\beta_\text{QED}^{(1)}(K)$ can be traced directly to the leading pole of the $\Gamma\!\left(\frac{5}{2}-\frac{x}{3}\right)$ factor in the integrand function $F_\text{QED}(x)$.
This gamma factor arises because it enters the leading $1/N_f$ computation via iteration of the 
 basic building block  of the one-bubble self-energy diagram, whose
 amplitude $\Pi_0$  is given by  
\begin{align} 
 \label{eq:one-bubble}
 \Pi_0 (\epsilon) \sim 2\frac{\Gamma^2\!\left(2-\frac{\epsilon}{2}\right)\Gamma\!\left(\frac{\epsilon}{2}\right)}{\Gamma\!\left(4-\epsilon\right)} \,,
\end{align}
regularized in $d=4-\epsilon$.
In the resummation, this amplitude typically enters the full beta-function as 
its inverse, $1/ \Pi_0$, and its argument is rescaled with the value of the $1/\epsilon$-pole\footnote{This is the case for diagrams containing one resummed gauge chain.}  \cite{Gracey:2018ame}. 
 
For the QED computation, the value of the $1/\epsilon$-pole in \eqref{eq:one-bubble} is $\frac23$.
Consequently, we expect the resummed $1/N_f$ beta-function to contain 
the factor
\begin{align}
 \label{eq:factor-beta}
 \Pi_0^{-1}\!\left( \frac 23 x \right) = 
 \frac{ 2^{2-\frac{2 x}{3}} \sin\!\left(\frac{\pi  x}{3}\right) \Gamma\!\left(\frac{5}{2}-\frac{x}{3}\right)}{\pi ^{3/2} (1-\frac{x}{3})} \,.
\end{align}
Indeed, this factor appears in the integrand function $F_\text{QED}(x)$, and governs the pole structure underlying the asymptotics of the perturbative expansion coefficients, and the structure of the Pad\'e approximations to the integrand function $F_\text{QED}(x)$.

Knowing this, one can devise {\it improved} expansions in which this $ \Pi_0^{-1}\!\left( \frac 23 x \right)$ factor is factored out, with only the remaining factors needing to be analyzed. Not surprisingly, this leads to noticeable improvement of the resulting Pad\'e approximations, and a much faster approach to the asymptotic behavior of the expansion coefficients.

The general idea of using Pad\'e approximants to study the behavior of beta-functions is significant for analyzing higher orders in the $1/N_f$ expansion, for which no closed-form resummation formula is currently known. In particular, these methods may allow us to access the leading-order pole structure at higher orders in the $1/N_f$ expansion, if enough coefficients can be extracted from the relevant diagrams. This information has quantitative implications for the stability, size, and structure of the asymptotically safe conformal window. 
One can imagine the following possible scenarios at higher orders in the $1/N_f$ expansion: 
\begin{itemize}
 \item[i)] A new singular structure may emerge closer to the origin, de facto, disconnecting the putative fixed point in \eqref{eq:npk}  from the 
 Gau\ss ian fixed point at the origin. 
The detailed structure of the new singularity would determine whether or not the theory remains UV safe to this order.
Alternatively if the radius of convergence of the series keeps shrinking as the order in $1/N_f$ increases, the UV fixed point could eventually disappear.  
\item[ii)] The current singular structure, and its location, could be further reinforced by higher-order corrections. This possibility is partially supported by the fact that the fermion self-energy amplitude is responsible for the singular structure of the theory. 
The order of the pole in the integrand might become stronger because  $n$ bubble chains appear in diagrams at the order $\mathcal{O}(1/N_f^n)$.  In this case the ultimate UV fate of the theory will depend on the character, sign, and strength of the reinforced singular structure.
\item[iii)] No further singularities emerge, or a new singular structure appears further away from the leading-order one.  This would be an indication that the putative fixed point in \eqref{eq:npk} is indeed physical. For example, the leading isolated pole, like the one that we will see appearing in QCD in the next section, is a candidate for this scenario since it is not due to the fermion self-energy amplitude.
\end{itemize}
Of course, even the ultimate confirmation of a non-perturbative zero in a generic beta-function away from the origin, is typically insufficient to establish the existence of a physical conformal field theory.  Other critical quantities such as the variation of the a-function or anomalous dimensions can potentially violate physical bounds \cite{Intriligator:2015xxa,Antipin:2017ebo}.

\section{Comparing different theories and their physics}
\label{sec:comparison}

\subsection{Large \texorpdfstring{$N_f$}{N_f}  QCD}
\label{sec:qcd}
The beta-function at order $1/N_f$  for an $SU(N_c)$ gauge theory was first calculated in  
\cite{Gracey:1996he}, and written in a closed integral form in \cite{Holdom:2010qs}. The result is: 
\begin{align}
 \label{eq:QCD}
 \beta_\text{QCD}(K) &= \frac{2K^2}{3} \left( 1 - \frac{11}{4 N_f } \frac{C_2(G)}{S_2(R)}\right)
 \\
  &  + \frac{K^2}{2N_f} \int_0^K \!\! \mathrm  dx \, F_\text{QCD} (x)+O\left(\frac{1}{N_f^2}\right)  \,.
  \notag
\end{align}
The integrand function is now
\begin{align}
 \label{eq:QCDF}
 F_\text{QCD}(x) &= \frac{2^{1-\frac{2 x}{3}} \sin\!\left(\frac{\pi  x}{3}\right) 
   \Gamma\!\left(\frac{5}{2}-\frac{x}{3}\right)}{27 \pi ^{3/2}  (x-3)^2 x\, \Gamma\!\left(3-\frac{x}{3}\right)} 
  \\
 & \quad \times 
  \Big[ \frac{C_2(G)}{S_2(R)} (4 x^4-42 x^3+288 x^2-1161 x)
 \notag \\
 &\qquad-4 \frac{d(G)}{ d(R)} (x - 3) (x + 3) (2 x  - 9) (2 x - 3 )\Big] \,.
 \notag
\end{align}
where $d(G)$, $d(R)$ are the dimensions of the group $G$ and the representation $R$, and similarly for the quadratic Casimirs $C_2$.
This result is very similar to that for QED and agrees with it in the Abelian limit $C_2(G)\to 0$ and $d(G)/d(R)\to1$.
The gamma factor $\Gamma\!\left(\frac52 - \frac{x}3\right)$ produces the same pole pattern as in QED.
However, a new isolated simple pole appears at $x_*=3$, leading to a smaller radius of convergence, $K_*=3$. This effect is purely due to the non-Abelian nature of the theory, as can be seen also from its residue:
\begin{align}
R_0 = \frac{1}{12} \frac{C_2(G)}{{S_2(R)}} \,, \qquad \text{at} \quad x_*=3 \,.
\end{align}
By an argument similar to that in \autoref{sec:origin}, we can identify this pole with the gluon bubble-loop rather than the fermion bubble-loop. Since this diagram does not appear iterated in a chain, it does not result in an entire series of poles.

\begin{figure}[t!]
\includegraphics[width=\linewidth]{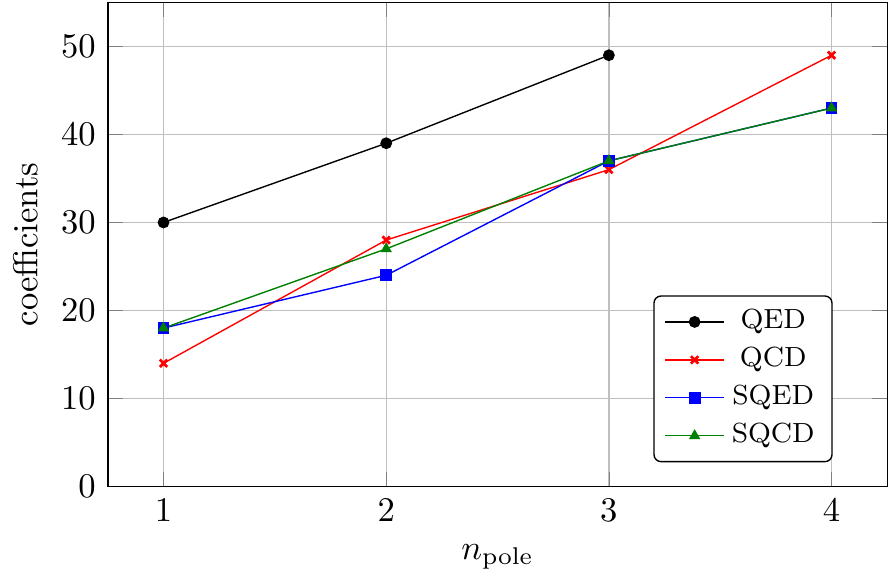}
\caption{Number of poles in each theory, ordered according to \autoref{tab:poles_residues}, 
versus the corresponding number of coefficients needed to resolve the pole.
We determine the number of coefficients as the minimal number of terms
needed to calculate the corresponding residue with a precision of $10^{-3}$. }
\label{fig:order_comparison}
\end{figure}

The simple pole of $F_\text{QCD}(x)$ at $x_*=3$ leads to a logarithmic behavior of $\beta^{(1)}_\text{QCD}(K)$ 
\begin{align}
\label{eq:log-qcd}
\beta^{(1)}_\text{QCD}(K)\sim &\frac{K^2}{24} \frac{C_2(G)}{S_2(R)} \ln\left(3-K\right)+\dots \,\, , \quad 
K\to 3 \,.
\end{align}
This implies that in order to obtain a zero of the beta-function in the large $N_f$ limit, we must approach a non-perturbative fixed point at \cite{Litim:2014uca,Antipin:2017ebo}
\begin{align}
K_*^\text{np}&=3-\exp\left[-16 \frac{S_2(R)}{C_2(G)} N _f  \right] \,.
\label{eq:npk-qcd}
\end{align}
Since the leading singularity for QCD is closer to the origin than for QED, fewer perturbative orders are required to resolve it using an asymptotic or Pad\'e analysis. For example, for $N_c=3$ and with fermions in the fundamental representation, the leading residue can be extracted with $\mathcal{O}(10^{-3})$ accuracy from the asymptotic expansion of the coefficients already at $\sim$14th order. Retaining up to 15th order in the expansion of the integrand, the Pad\'e approximant $\mathcal P^{[7,7]}(x)$ gives a good reconstruction of the integrand within the radius of convergence. A  similar analysis can be carried out for the other poles in the integrand: the results for the theories considered here are summarized in \autoref{fig:order_comparison}. Note that our result for QED is compatible with \cite{Shrock:2013cca} where an expansion up to the 24th order was not sufficient to find a stable zero in the beta-function.

Our results across the various theories indicate that the main factor determining the number of coefficients needed to resolve a given pole is the distance of the latter to the origin\footnote{A simple rescaling of the couplings does not change the number of coefficients needed.}.
As discussed in the end of \autoref{sec:origin}, the behavior near the leading pole is associated with the amplitude factor $\Pi_0^{-1}$, so we expect a similar relation between the number of coefficients and the location of the poles at the next orders in the $1/N_f$ expansion. Furthermore, since no closed form re-summed perturbative expressions are known at higher order in $1/N_f$, this motivates the importance of a similar Pad\'e analysis of the perturbative series at $1/N_f^2$.

\subsection{Large \texorpdfstring{$N_f$}{N_f} Supersymmetric Results}
In this section we review and discuss the results obtained in \cite{Ferreira:1997bi,Ferreira:1997rc} for large-$N_f$ $\mathcal{N}=1$ supersymmetric QED and QCD. The results are obtained in dimensional reduction (DRED) in $d=4-2\epsilon$. For SQED one finds:
\begin{align}
 \label{SQED}
 \beta_\text{SQED}(K) &= K^2 + \frac{K^2}{2N_f}  \int_0^K \!\! \mathrm  dx \, F_\text{SQED}(x)+\dots \,,
 \notag \\
 F_\text{SQED}(x)
 &=\frac{2^{3-x} (1-x) \sin\!\left(\frac{\pi  x}{2}\right) \Gamma\!\left(\frac{3}{2}-\frac{x}{2}\right)}{\pi ^{3/2} x\,\Gamma\!\left(2-\frac{x}{2}\right)} \,.
\end{align}
While for SQCD:  
\begin{align}
 \label{eq:SQCD}
 \beta_\text{SQCD}(K) &= K^2 \left( 1 - \frac{1}{N_f} \frac{C_2(G)}{S_2(R)} \right)
 \notag \\
 & \quad+ \frac{K^2}{2N_f}  \int_0^K \!\! \mathrm dx \,F_\text{SQCD}(x) +\dots \notag  \,,
\\ 
F_\text{SQCD}(x) &= \frac{2^{2-x} \sin\!\left(\frac{\pi  x}{2}\right) \Gamma\!\left(\frac{3}{2}-\frac{x}{2}\right)}
  {\pi ^{3/2} x\,\Gamma\!\left(2-\frac{x}{2}\right)} 
  \notag \\
  &\quad \times \left[ 2(1-x) \frac{d(G)}{d(R)}  + \frac{C_2(G)}{S_2(R)} \right]\,.
\end{align}
Notice again that the integrand expression for QCD in \eqref{eq:QCDF} agrees with the one for QED in \eqref{eq:QEDF} the Abelian limit.
For each of the SUSY beta-functions in \eqref{SQED} and \eqref{eq:SQCD}, the integrand function $F(x)$ has its first singularity as a simple pole at $x=3$ with a negative residue\footnote{In our convention the coupling $K$ is twice the coupling defined in the original works, where the pole appeared at $x=3/2$.}, see \autoref{tab:poles_residues}. For example, this opposite sign explains the different behavior of the beta-function near the first singularity, as shown in \autoref{fig:qed-sqed} for QED and SQED. Due to this negative sign for SQED and SQCD, the associated logarithmic singularity of the beta-function cannot provide a cancellation between the first two orders in the large-$N_f$ expansion \eqref{eq:cancel}, and therefore no non-perturbative fixed point arises. 

It is interesting to note that this conclusion holds also in the NSVZ scheme, which can be related to DRED in by an order-by-order coupling redefinition \cite{Jack:1996cn}, see also \cite{Kataev:2013csa} for details on such a relation. The well known NSVZ beta-function (see \cite{Shifman:1996iy} for a recent discussion) is 
\begin{align}
\label{eq:beta-nsvz}
\beta^\text{NSVZ}(g) = - \frac{g^3}{(4\pi)^2} \frac{3C_2(G) - N_f S_2(R) (1 - \gamma^\text{NSVZ}(g) )}{1 - \frac{2g^2}{(4\pi)^2} C_2(G)}\,.
\end{align}
It admits a  zero where the anomalous dimension takes the value
\begin{align}
\gamma(g^*) = 1 - \frac{1}{N_f} \frac{3C_2(G)}{S_2(R)} \,.
\end{align}
We dropped the NSVZ label, as this quantity is scheme independent at the alleged fixed point. In the limit $N_f \gg N_c$, the theory has lost asymptotic freedom and therefore such a fixed point has to develop in the UV. However due to the violation of the   $a$-theorem \cite{Intriligator:2015xxa,Martin:2000cr,Bajc:2017xwx}  it is disconnected from the IR Gau\ss ian fixed point.  The absence of an UV fixed point smoothly connected to the origin agrees with the large-$N_f$ result in the DRED scheme, where, in fact, no UV fixed point is seen to complete the Gau\ss ian. 

\begin{table}[t]
\centering
    \caption{Poles and first residue of the $1/N_f$ resummed integrand function $F(x)$ for each theory considered.}
    \label{tab:poles_residues}
    \begin{tabular}{ | l | l | l |}
    \hline
    & Poles & $ \text{R}_0 $  \\ \hline
    QED & $x_n = 3n + \frac{15}{2}$ \,, \qquad $n\geq 0$  &  $\frac{28}{45 \pi^2}$ \\[1ex] 
    \hline
    \midrule
 \multirow{2}{*}{QCD} & $x_0 = 3$ &    \multirow{2}{*}{$\frac{1}{12}\frac{C_2(G)}{{S_2(R)}}$}   \\
    	&  $x_n = 3(n-1)+\frac{15}{2}\,,\quad n\geq1$  &  \\[1ex] 
    \hline
    SQED & $x_n = 3 + 2n $ \,, \qquad $n\geq 0$ & $- \frac{4}{3\pi^2}$   \\[1ex] 
    \hline
    SQCD & $x_n = 3 + 2n$ \,, \qquad $n\geq 0$ & $- \frac{4}{3\pi^2}{ \left( \frac{d(G)}{d(R)} - \frac{1}{4} \frac{C_2(G)}{S_2(R)} \right)}$ \\[1ex] 
    \hline
    \end{tabular}
\end{table}

\section{Conclusions}
\label{sec:conclusions}
Our analysis of the convergence properties of the leading  $1/N_f$ behavior of QED, QCD and their supersymmetric cousins has revealed several interesting features. We observe the emergence of a common analytic structure stemming from the leading $1/N_f$ corrections, with the important difference that the coefficient of the logarithmic branch singularity is positive for QED and QCD, but it switches sign for their supersymmetric counterparts. The sign plays a crucial r\^ole when considering the UV fate of these theories. For example for the supersymmetric case it implies that the theories are not fundamental, in agreement with other non-perturbative analyses. 

We have demonstrated, by direct comparison with the full result, that the analysis of the large-order behavior of the 't Hooft coupling expansion is able to identify the location and nature of the leading logarithmic singularities, including the overall sign and magnitude of their coefficients. This suggests that a large-order analysis can be used in the near future to tackle the next-to-leading order in the $1/N_f$ expansion, in the absence of a closed-form result. These corrections are crucial to test the singular structure of the leading $1/N_f$ result, with important consequences for the UV fate of these theories. 

\vspace{.3cm}
\noindent {\bf Acknowledgements} 
This work is partially supported by the Danish National Research Foundation grant DNRF:90, and  is based upon work supported by the U.S. Department of Energy, Office of Science, Office of High Energy Physics under Award Number DE-SC0010339. 

\appendix

\section{Darboux's Theorem and Large-order Behavior of Expansion Coefficients}
\label{app:Darboux}

Darboux's theorem says that for a convergent series expansion, the behavior of the expansion in the vicinity of a nearby singularity is determined by the large-order growth of the expansion coefficients about another point (say $z=0$) \cite{Fisher,Henrici,Guttmann}. 
For example, suppose a function $f(z)$ has the following expansion in the vicinity of a point $z_0$:
\begin{align}
f(z)\sim  \phi(z)\, \left(1-\frac{z}{z_0}\right)^{-p}+\psi(z) \,, \qquad z\to z_0 \,,
\label{eq:darboux1}
\end{align}
where $\phi(z)$ and $\psi(z)$ are analytic near $z_0$. 
Then the Taylor expansion coefficients of $f(z)$ near the origin have large-order growth
\begin{align}
b_n \sim{}& \frac{1}{z_0^n}\begin{pmatrix}
n+p-1\\ n
\end{pmatrix} \left[ \phi(z_0)- 
\frac{(p-1)\, z_0\, \phi^\prime(z_0)}{(n+p-1)} \right.
\nonumber\\[1ex]
&
\left. +
\frac{(p-1)(p-2)\, z_0^2\, \phi^{\prime\prime}(z_0)}{2! (n+p-1)(n+p-2)}\, -\dots \right] \,,
\label{eq:darboux2}
\end{align}
This argument can be run in reverse, so that an analysis of the large-order behavior of the coefficients $b_n$ enables us first to determine the radius of convergence $z_0$, and then also the nature $p$ of the singularity; for example,  a pole, or a branch-cut, and what type of branch-cut. The overall factor determines $\phi(z_0)$, and further sub-leading information determines higher orders of the expansion of $\phi(z)$ about $z_0$.
If the singularity is logarithmic,
\begin{align}
f(z)\sim  \phi(z)\, \ln\left(1-\frac{z}{z_0}\right)+\psi(z) \,, \qquad z\to z_0 \,,
\label{eq:darboux3}
\end{align}
where $\phi(z)$ and $\psi(z)$ are analytic near $z_0$, 
then the Taylor expansion coefficients of $f(z)$ near the origin have large-order growth
\begin{align}
b_n\sim \frac{1}{z_0^n} \frac{1}{n} \left[\phi(z_0) - \frac{z_0\, \phi^\prime(z_0)}{(n-1)} +\frac{z_0^2\, \phi^{\prime\prime}(z_0)}{(n-1)(n-2)} -\dots \right]\,,
\label{eq:darboux4}
\end{align}
Once again, the large-order behavior of the convergent expansion coefficients determines the nature of the singularity, and the fluctuations about it.

\section{Obtaining the Numerical Coefficients of the Beta Functions}
\label{BAppendix}

The $\mathcal{O}\!\left(1/N_f \right)$  beta functions discussed in this work are known in their resummed form, so we can re-expand them to 60 orders and perform a Pad\'e analysis. The motivation for this is to obtain an estimate of how many perturbative terms are required in order to identify both the location and nature of the leading singularity, with a view towards a direct perturbative computation of the $\mathcal{O}(1/N_f^2)$ beta functions, for which no resummed version is currently known. It is not {\it a priori} clear whether one might need 10 terms, or several hundred. Our work has shown that at $\mathcal{O}(1/N_f)$ roughly 30-40 perturbative terms are required, due to the lower order oscillatory behavior, which we have associated with the appearance of the amplitude factor in \eqref{eq:one-bubble}. Since these factors appear also in an $\mathcal{O}(1/N_f^2)$ computation, we expect that at least the same number of perturbative coefficients would be necessary in such a computation at a given higher order in the $1/N_f$ expansion.
We now comment briefly on the steps required to make such a computation feasible at higher orders in the $1/N_f$ expansion. To this end, we first describe how the $\mathcal{O}\!\left(1/N_f \right)$ perturbative expansion is obtained in the diagrammatic approach.

The diagrams that contribute to the beta function at the order $1/N_f$ in QED and QCD are displayed, e.g., in \cite{Antipin:2018zdg}.
For the contraction of the diagrams  one can use the Mathematica package \textit{FeynCalc} \cite{Shtabovenko:2016sxi},
which performs the trace over Lorentz and Dirac indices in $d$ dimensions.
Complicated diagrams can be traced with the symbolic manipulation system \textit{FORM} \cite{Vermaseren:2000nd,Kuipers:2012rf},
as well as the Mathematica package \textit{FormTracer} \cite{Cyrol:2016zqb}.
The contracted diagrams  can be evaluated with standard multi-loop techniques along the lines of
\cite{Broadhurst:1996ur,Bierenbaum:2003ud,Grozin:2003ak,Grozin:2012xi}.
The diagrams contain fully dressed gauge propagators and thus one has to apply reduction formulas that only hit non-dressed propagators.

This procedure can be extended to higher orders in the $1/N_f$ expansion, to determine the perturbative coefficients  up to arbitrary order in the coupling but subleading in $1/N_f$. At higher order in $1/N_f$, the loop-order and the number of dressed propagators increases. Thus, at higher order in $1/N_f$, we expect that more efficient reduction formulae will be required.

Another complication at higher order in $1/N_f$ is that at any order the integrated diagrams contain gamma and hypergeometric functions, which need to be expanded in $\epsilon=d-4$. For the $n$-th loop coefficient we need to expand these functions up to $\epsilon^{n-1}$. These expansions are slow, and at  higher order in $1/N_f$ there will be more such factors to expand.
Analytic expansions are only known for specific cases or only to low order.
Thus in our $\mathcal{O}\!\left(1/N_f \right)$  computation, we used numerical expansion methods, in particular the package \textit{NumExp} \cite{Huang:2012qz}.
 This numerical precision will need to be balanced with the precision required for the subsequent Pad\'e analysis.

\bibliography{largeN}

\begin{thebibliography}{42}%
\makeatletter
\providecommand \@ifxundefined [1]{%
 \@ifx{#1\undefined}
}%
\providecommand \@ifnum [1]{%
 \ifnum #1\expandafter \@firstoftwo
 \else \expandafter \@secondoftwo
 \fi
}%
\providecommand \@ifx [1]{%
 \ifx #1\expandafter \@firstoftwo
 \else \expandafter \@secondoftwo
 \fi
}%
\providecommand \natexlab [1]{#1}%
\providecommand \enquote  [1]{``#1''}%
\providecommand \bibnamefont  [1]{#1}%
\providecommand \bibfnamefont [1]{#1}%
\providecommand \citenamefont [1]{#1}%
\providecommand \href@noop [0]{\@secondoftwo}%
\providecommand \href [0]{\begingroup \@sanitize@url \@href}%
\providecommand \@href[1]{\@@startlink{#1}\@@href}%
\providecommand \@@href[1]{\endgroup#1\@@endlink}%
\providecommand \@sanitize@url [0]{\catcode `\\12\catcode `\$12\catcode
  `\&12\catcode `\#12\catcode `\^12\catcode `\_12\catcode `\%12\relax}%
\providecommand \@@startlink[1]{}%
\providecommand \@@endlink[0]{}%
\providecommand \url  [0]{\begingroup\@sanitize@url \@url }%
\providecommand \@url [1]{\endgroup\@href {#1}{\urlprefix }}%
\providecommand \urlprefix  [0]{URL }%
\providecommand \Eprint [0]{\href }%
\providecommand \doibase [0]{http://dx.doi.org/}%
\providecommand \selectlanguage [0]{\@gobble}%
\providecommand \bibinfo  [0]{\@secondoftwo}%
\providecommand \bibfield  [0]{\@secondoftwo}%
\providecommand \translation [1]{[#1]}%
\providecommand \BibitemOpen [0]{}%
\providecommand \bibitemStop [0]{}%
\providecommand \bibitemNoStop [0]{.\EOS\space}%
\providecommand \EOS [0]{\spacefactor3000\relax}%
\providecommand \BibitemShut  [1]{\csname bibitem#1\endcsname}%
\let\auto@bib@innerbib\@empty
\bibitem [{\citenamefont {Litim}\ and\ \citenamefont
  {Sannino}(2014)}]{Litim:2014uca}%
  \BibitemOpen
  \bibfield  {author} {\bibinfo {author} {\bibfnamefont {D.~F.}\ \bibnamefont
  {Litim}}\ and\ \bibinfo {author} {\bibfnamefont {F.}~\bibnamefont
  {Sannino}},\ }\href {\doibase 10.1007/JHEP12(2014)178} {\bibfield  {journal}
  {\bibinfo  {journal} {JHEP}\ }\textbf {\bibinfo {volume} {12}},\ \bibinfo
  {pages} {178} (\bibinfo {year} {2014})},\ \Eprint
  {http://arxiv.org/abs/1406.2337} {arXiv:1406.2337 [hep-th]} \BibitemShut
  {NoStop}%
\bibitem [{\citenamefont {Leino}\ \emph {et~al.}()\citenamefont {Leino},
  \citenamefont {Rindlisbacher}, \citenamefont {Rummukainen}, \citenamefont
  {Sannino},\ and\ \citenamefont {Tuominen}}]{LatticeNf}%
  \BibitemOpen
  \bibfield  {author} {\bibinfo {author} {\bibfnamefont {V.}~\bibnamefont
  {Leino}}, \bibinfo {author} {\bibfnamefont {T.}~\bibnamefont
  {Rindlisbacher}}, \bibinfo {author} {\bibfnamefont {K.}~\bibnamefont
  {Rummukainen}}, \bibinfo {author} {\bibfnamefont {F.}~\bibnamefont
  {Sannino}}, \ and\ \bibinfo {author} {\bibfnamefont {K.}~\bibnamefont
  {Tuominen}},\ }\href@noop {} {\ }\bibinfo {note} {To appear}\BibitemShut
  {NoStop}%
\bibitem [{\citenamefont {Palanques-Mestre}\ and\ \citenamefont
  {Pascual}(1984)}]{PalanquesMestre:1983zy}%
  \BibitemOpen
  \bibfield  {author} {\bibinfo {author} {\bibfnamefont {A.}~\bibnamefont
  {Palanques-Mestre}}\ and\ \bibinfo {author} {\bibfnamefont {P.}~\bibnamefont
  {Pascual}},\ }\href {\doibase 10.1007/BF01212398} {\bibfield  {journal}
  {\bibinfo  {journal} {Commun. Math. Phys.}\ }\textbf {\bibinfo {volume}
  {95}},\ \bibinfo {pages} {277} (\bibinfo {year} {1984})}\BibitemShut
  {NoStop}%
\bibitem [{\citenamefont {Gracey}(1996)}]{Gracey:1996he}%
  \BibitemOpen
  \bibfield  {author} {\bibinfo {author} {\bibfnamefont {J.~A.}\ \bibnamefont
  {Gracey}},\ }\href {\doibase 10.1016/0370-2693(96)00105-0} {\bibfield
  {journal} {\bibinfo  {journal} {Phys. Lett.}\ }\textbf {\bibinfo {volume}
  {B373}},\ \bibinfo {pages} {178} (\bibinfo {year} {1996})},\ \Eprint
  {http://arxiv.org/abs/hep-ph/9602214} {arXiv:hep-ph/9602214 [hep-ph]}
  \BibitemShut {NoStop}%
\bibitem [{\citenamefont {Holdom}(2011)}]{Holdom:2010qs}%
  \BibitemOpen
  \bibfield  {author} {\bibinfo {author} {\bibfnamefont {B.}~\bibnamefont
  {Holdom}},\ }\href {\doibase 10.1016/j.physletb.2010.09.037} {\bibfield
  {journal} {\bibinfo  {journal} {Phys. Lett.}\ }\textbf {\bibinfo {volume}
  {B694}},\ \bibinfo {pages} {74} (\bibinfo {year} {2011})},\ \Eprint
  {http://arxiv.org/abs/1006.2119} {arXiv:1006.2119 [hep-ph]} \BibitemShut
  {NoStop}%
\bibitem [{\citenamefont {Gracey}(2019)}]{Gracey:2018ame}%
  \BibitemOpen
  \bibfield  {author} {\bibinfo {author} {\bibfnamefont {J.~A.}\ \bibnamefont
  {Gracey}},\ }\href {\doibase 10.1142/S0217751X18300326} {\bibfield  {journal}
  {\bibinfo  {journal} {Int. J. Mod. Phys.}\ }\textbf {\bibinfo {volume}
  {A33}},\ \bibinfo {pages} {1830032} (\bibinfo {year} {2019})},\ \Eprint
  {http://arxiv.org/abs/1812.05368} {arXiv:1812.05368 [hep-th]} \BibitemShut
  {NoStop}%
\bibitem [{\citenamefont {Mann}\ \emph {et~al.}(2017)\citenamefont {Mann},
  \citenamefont {Meffe}, \citenamefont {Sannino}, \citenamefont {Steele},
  \citenamefont {Wang},\ and\ \citenamefont {Zhang}}]{Mann:2017wzh}%
  \BibitemOpen
  \bibfield  {author} {\bibinfo {author} {\bibfnamefont {R.}~\bibnamefont
  {Mann}}, \bibinfo {author} {\bibfnamefont {J.}~\bibnamefont {Meffe}},
  \bibinfo {author} {\bibfnamefont {F.}~\bibnamefont {Sannino}}, \bibinfo
  {author} {\bibfnamefont {T.}~\bibnamefont {Steele}}, \bibinfo {author}
  {\bibfnamefont {Z.-W.}\ \bibnamefont {Wang}}, \ and\ \bibinfo {author}
  {\bibfnamefont {C.}~\bibnamefont {Zhang}},\ }\href {\doibase
  10.1103/PhysRevLett.119.261802} {\bibfield  {journal} {\bibinfo  {journal}
  {Phys. Rev. Lett.}\ }\textbf {\bibinfo {volume} {119}},\ \bibinfo {pages}
  {261802} (\bibinfo {year} {2017})},\ \Eprint
  {http://arxiv.org/abs/1707.02942} {arXiv:1707.02942 [hep-ph]} \BibitemShut
  {NoStop}%
\bibitem [{\citenamefont {Pelaggi}\ \emph {et~al.}(2018)\citenamefont
  {Pelaggi}, \citenamefont {Plascencia}, \citenamefont {Salvio}, \citenamefont
  {Sannino}, \citenamefont {Smirnov},\ and\ \citenamefont
  {Strumia}}]{Pelaggi:2017abg}%
  \BibitemOpen
  \bibfield  {author} {\bibinfo {author} {\bibfnamefont {G.~M.}\ \bibnamefont
  {Pelaggi}}, \bibinfo {author} {\bibfnamefont {A.~D.}\ \bibnamefont
  {Plascencia}}, \bibinfo {author} {\bibfnamefont {A.}~\bibnamefont {Salvio}},
  \bibinfo {author} {\bibfnamefont {F.}~\bibnamefont {Sannino}}, \bibinfo
  {author} {\bibfnamefont {J.}~\bibnamefont {Smirnov}}, \ and\ \bibinfo
  {author} {\bibfnamefont {A.}~\bibnamefont {Strumia}},\ }\href {\doibase
  10.1103/PhysRevD.97.095013} {\bibfield  {journal} {\bibinfo  {journal} {Phys.
  Rev.}\ }\textbf {\bibinfo {volume} {D97}},\ \bibinfo {pages} {095013}
  (\bibinfo {year} {2018})},\ \Eprint {http://arxiv.org/abs/1708.00437}
  {arXiv:1708.00437 [hep-ph]} \BibitemShut {NoStop}%
\bibitem [{\citenamefont {Antipin}\ \emph {et~al.}(2018)\citenamefont
  {Antipin}, \citenamefont {Dondi}, \citenamefont {Sannino}, \citenamefont
  {Thomsen},\ and\ \citenamefont {Wang}}]{Antipin:2018zdg}%
  \BibitemOpen
  \bibfield  {author} {\bibinfo {author} {\bibfnamefont {O.}~\bibnamefont
  {Antipin}}, \bibinfo {author} {\bibfnamefont {N.~A.}\ \bibnamefont {Dondi}},
  \bibinfo {author} {\bibfnamefont {F.}~\bibnamefont {Sannino}}, \bibinfo
  {author} {\bibfnamefont {A.~E.}\ \bibnamefont {Thomsen}}, \ and\ \bibinfo
  {author} {\bibfnamefont {Z.-W.}\ \bibnamefont {Wang}},\ }\href {\doibase
  10.1103/PhysRevD.98.016003} {\bibfield  {journal} {\bibinfo  {journal} {Phys.
  Rev.}\ }\textbf {\bibinfo {volume} {D98}},\ \bibinfo {pages} {016003}
  (\bibinfo {year} {2018})},\ \Eprint {http://arxiv.org/abs/1803.09770}
  {arXiv:1803.09770 [hep-ph]} \BibitemShut {NoStop}%
\bibitem [{\citenamefont {Alanne}\ and\ \citenamefont
  {Blasi}(2018{\natexlab{a}})}]{Alanne:2018ene}%
  \BibitemOpen
  \bibfield  {author} {\bibinfo {author} {\bibfnamefont {T.}~\bibnamefont
  {Alanne}}\ and\ \bibinfo {author} {\bibfnamefont {S.}~\bibnamefont {Blasi}},\
  }\href {\doibase 10.1007/JHEP08(2018)081, 10.1007/JHEP09(2018)165} {\bibfield
   {journal} {\bibinfo  {journal} {JHEP}\ }\textbf {\bibinfo {volume} {08}},\
  \bibinfo {pages} {081} (\bibinfo {year} {2018}{\natexlab{a}})},\ \bibinfo
  {note} {[Erratum: JHEP09,165(2018)]},\ \Eprint
  {http://arxiv.org/abs/1806.06954} {arXiv:1806.06954 [hep-ph]} \BibitemShut
  {NoStop}%
\bibitem [{\citenamefont {Alanne}\ and\ \citenamefont
  {Blasi}(2018{\natexlab{b}})}]{Alanne:2018csn}%
  \BibitemOpen
  \bibfield  {author} {\bibinfo {author} {\bibfnamefont {T.}~\bibnamefont
  {Alanne}}\ and\ \bibinfo {author} {\bibfnamefont {S.}~\bibnamefont {Blasi}},\
  }\href {\doibase 10.1103/PhysRevD.98.116004} {\bibfield  {journal} {\bibinfo
  {journal} {Phys. Rev.}\ }\textbf {\bibinfo {volume} {D98}},\ \bibinfo {pages}
  {116004} (\bibinfo {year} {2018}{\natexlab{b}})},\ \Eprint
  {http://arxiv.org/abs/1808.03252} {arXiv:1808.03252 [hep-ph]} \BibitemShut
  {NoStop}%
\bibitem [{\citenamefont {Kowalska}\ and\ \citenamefont
  {Sessolo}(2018)}]{Kowalska:2017pkt}%
  \BibitemOpen
  \bibfield  {author} {\bibinfo {author} {\bibfnamefont {K.}~\bibnamefont
  {Kowalska}}\ and\ \bibinfo {author} {\bibfnamefont {E.~M.}\ \bibnamefont
  {Sessolo}},\ }\href {\doibase 10.1007/JHEP04(2018)027} {\bibfield  {journal}
  {\bibinfo  {journal} {JHEP}\ }\textbf {\bibinfo {volume} {04}},\ \bibinfo
  {pages} {027} (\bibinfo {year} {2018})},\ \Eprint
  {http://arxiv.org/abs/1712.06859} {arXiv:1712.06859 [hep-ph]} \BibitemShut
  {NoStop}%
\bibitem [{\citenamefont {Antipin}\ and\ \citenamefont
  {Sannino}(2018)}]{Antipin:2017ebo}%
  \BibitemOpen
  \bibfield  {author} {\bibinfo {author} {\bibfnamefont {O.}~\bibnamefont
  {Antipin}}\ and\ \bibinfo {author} {\bibfnamefont {F.}~\bibnamefont
  {Sannino}},\ }\href {\doibase 10.1103/PhysRevD.97.116007} {\bibfield
  {journal} {\bibinfo  {journal} {Phys. Rev.}\ }\textbf {\bibinfo {volume}
  {D97}},\ \bibinfo {pages} {116007} (\bibinfo {year} {2018})},\ \Eprint
  {http://arxiv.org/abs/1709.02354} {arXiv:1709.02354 [hep-ph]} \BibitemShut
  {NoStop}%
\bibitem [{\citenamefont {Pica}\ and\ \citenamefont
  {Sannino}(2011)}]{Pica:2010xq}%
  \BibitemOpen
  \bibfield  {author} {\bibinfo {author} {\bibfnamefont {C.}~\bibnamefont
  {Pica}}\ and\ \bibinfo {author} {\bibfnamefont {F.}~\bibnamefont {Sannino}},\
  }\href {\doibase 10.1103/PhysRevD.83.035013} {\bibfield  {journal} {\bibinfo
  {journal} {Phys. Rev.}\ }\textbf {\bibinfo {volume} {D83}},\ \bibinfo {pages}
  {035013} (\bibinfo {year} {2011})},\ \Eprint {http://arxiv.org/abs/1011.5917}
  {arXiv:1011.5917 [hep-ph]} \BibitemShut {NoStop}%
\bibitem [{\citenamefont {Sannino}\ and\ \citenamefont
  {Tuominen}(2005)}]{Sannino:2004qp}%
  \BibitemOpen
  \bibfield  {author} {\bibinfo {author} {\bibfnamefont {F.}~\bibnamefont
  {Sannino}}\ and\ \bibinfo {author} {\bibfnamefont {K.}~\bibnamefont
  {Tuominen}},\ }\href {\doibase 10.1103/PhysRevD.71.051901} {\bibfield
  {journal} {\bibinfo  {journal} {Phys. Rev.}\ }\textbf {\bibinfo {volume}
  {D71}},\ \bibinfo {pages} {051901} (\bibinfo {year} {2005})},\ \Eprint
  {http://arxiv.org/abs/hep-ph/0405209} {arXiv:hep-ph/0405209 [hep-ph]}
  \BibitemShut {NoStop}%
\bibitem [{\citenamefont {Dietrich}\ and\ \citenamefont
  {Sannino}(2007)}]{Dietrich:2006cm}%
  \BibitemOpen
  \bibfield  {author} {\bibinfo {author} {\bibfnamefont {D.~D.}\ \bibnamefont
  {Dietrich}}\ and\ \bibinfo {author} {\bibfnamefont {F.}~\bibnamefont
  {Sannino}},\ }\href {\doibase 10.1103/PhysRevD.75.085018} {\bibfield
  {journal} {\bibinfo  {journal} {Phys. Rev.}\ }\textbf {\bibinfo {volume}
  {D75}},\ \bibinfo {pages} {085018} (\bibinfo {year} {2007})},\ \Eprint
  {http://arxiv.org/abs/hep-ph/0611341} {arXiv:hep-ph/0611341 [hep-ph]}
  \BibitemShut {NoStop}%
\bibitem [{\citenamefont {Caswell}(1974)}]{Caswell:1974gg}%
  \BibitemOpen
  \bibfield  {author} {\bibinfo {author} {\bibfnamefont {W.~E.}\ \bibnamefont
  {Caswell}},\ }\href {\doibase 10.1103/PhysRevLett.33.244} {\bibfield
  {journal} {\bibinfo  {journal} {Phys. Rev. Lett.}\ }\textbf {\bibinfo
  {volume} {33}},\ \bibinfo {pages} {244} (\bibinfo {year} {1974})}\BibitemShut
  {NoStop}%
\bibitem [{\citenamefont {Kreimer}(2000)}]{Kreimer}%
  \BibitemOpen
  \bibfield  {author} {\bibinfo {author} {\bibfnamefont {D.}~\bibnamefont
  {Kreimer}},\ }\href@noop {} {\emph {\bibinfo {title} {{Knots and Feynman
  Diagrams}}}}\ (\bibinfo  {publisher} {Cambridge University Press},\ \bibinfo
  {year} {2000})\BibitemShut {NoStop}%
\bibitem [{\citenamefont {Weinzierl}()}]{Weinzierl:2003jx}%
  \BibitemOpen
  \bibfield  {author} {\bibinfo {author} {\bibfnamefont {S.}~\bibnamefont
  {Weinzierl}},\ }in\ \href {\doibase 10.1007/978-3-540-30308-4_15} {\emph
  {\bibinfo {booktitle} {{Proceedings, Les Houches School of Physics: Frontiers
  in Number Theory, Physics and Geometry II: Les Houches, 2003}}}},\ \Eprint
  {http://arxiv.org/abs/hep-th/0305260} {arXiv:hep-th/0305260 [hep-th]}
  \BibitemShut {NoStop}%
\bibitem [{\citenamefont {Fisher}(1974)}]{Fisher}%
  \BibitemOpen
  \bibfield  {author} {\bibinfo {author} {\bibfnamefont {M.~E.}\ \bibnamefont
  {Fisher}},\ }\href {\doibase 10.1216/RMJ-1974-4-2-181} {\bibfield  {journal}
  {\bibinfo  {journal} {Rocky Mount. J. Math.}\ }\textbf {\bibinfo {volume}
  {4}},\ \bibinfo {pages} {181} (\bibinfo {year} {1974})}\BibitemShut {NoStop}%
\bibitem [{\citenamefont {Guttmann}(2016)}]{Guttmann}%
  \BibitemOpen
  \bibfield  {author} {\bibinfo {author} {\bibfnamefont {A.~J.}\ \bibnamefont
  {Guttmann}},\ }\href {\doibase 10.1088/1751-8113/49/41/415002} {\bibfield
  {journal} {\bibinfo  {journal} {Journal of Physics A: Mathematical and
  Theoretical}\ }\textbf {\bibinfo {volume} {49}},\ \bibinfo {pages} {415002}
  (\bibinfo {year} {2016})}\BibitemShut {NoStop}%
\bibitem [{\citenamefont {Henrici}(1977)}]{Henrici}%
  \BibitemOpen
  \bibfield  {author} {\bibinfo {author} {\bibfnamefont {P.}~\bibnamefont
  {Henrici}},\ }\href@noop {} {\emph {\bibinfo {title} {{Applied and
  Computational Complex Analysis}}}}\ (\bibinfo  {publisher} {Wiley},\ \bibinfo
  {year} {1977})\BibitemShut {NoStop}%
\bibitem [{\citenamefont {Bender}\ and\ \citenamefont {Orszag}(1999)}]{Bender}%
  \BibitemOpen
  \bibfield  {author} {\bibinfo {author} {\bibfnamefont {C.~M.}\ \bibnamefont
  {Bender}}\ and\ \bibinfo {author} {\bibfnamefont {S.~A.}\ \bibnamefont
  {Orszag}},\ }\href@noop {} {\emph {\bibinfo {title} {{Advanced Mathematical
  Methods for Scientists and Engineers}}}}\ (\bibinfo  {publisher} {Springer},\
  \bibinfo {year} {1999})\BibitemShut {NoStop}%
\bibitem [{\citenamefont {Baker}\ and\ \citenamefont
  {Graves-Morris}(1996)}]{Baker}%
  \BibitemOpen
  \bibfield  {author} {\bibinfo {author} {\bibfnamefont {G.~A.}\ \bibnamefont
  {Baker}}\ and\ \bibinfo {author} {\bibfnamefont {P.}~\bibnamefont
  {Graves-Morris}},\ }\href {\doibase 10.1017/CBO9780511530074} {\emph
  {\bibinfo {title} {Pade Approximants}}},\ \bibinfo {edition} {2nd}\ ed.,\
  Encyclopedia of Mathematics and its Applications\ (\bibinfo  {publisher}
  {Cambridge University Press},\ \bibinfo {year} {1996})\BibitemShut {NoStop}%
\bibitem [{\citenamefont {Intriligator}\ and\ \citenamefont
  {Sannino}(2015)}]{Intriligator:2015xxa}%
  \BibitemOpen
  \bibfield  {author} {\bibinfo {author} {\bibfnamefont {K.}~\bibnamefont
  {Intriligator}}\ and\ \bibinfo {author} {\bibfnamefont {F.}~\bibnamefont
  {Sannino}},\ }\href {\doibase 10.1007/JHEP11(2015)023} {\bibfield  {journal}
  {\bibinfo  {journal} {JHEP}\ }\textbf {\bibinfo {volume} {11}},\ \bibinfo
  {pages} {023} (\bibinfo {year} {2015})},\ \Eprint
  {http://arxiv.org/abs/1508.07411} {arXiv:1508.07411 [hep-th]} \BibitemShut
  {NoStop}%
\bibitem [{\citenamefont {Shrock}(2014)}]{Shrock:2013cca}%
  \BibitemOpen
  \bibfield  {author} {\bibinfo {author} {\bibfnamefont {R.}~\bibnamefont
  {Shrock}},\ }\href {\doibase 10.1103/PhysRevD.89.045019} {\bibfield
  {journal} {\bibinfo  {journal} {Phys. Rev.}\ }\textbf {\bibinfo {volume}
  {D89}},\ \bibinfo {pages} {045019} (\bibinfo {year} {2014})},\ \Eprint
  {http://arxiv.org/abs/1311.5268} {arXiv:1311.5268 [hep-th]} \BibitemShut
  {NoStop}%
\bibitem [{\citenamefont {Ferreira}\ \emph
  {et~al.}(1997{\natexlab{a}})\citenamefont {Ferreira}, \citenamefont {Jack},
  \citenamefont {Jones},\ and\ \citenamefont {North}}]{Ferreira:1997bi}%
  \BibitemOpen
  \bibfield  {author} {\bibinfo {author} {\bibfnamefont {P.~M.}\ \bibnamefont
  {Ferreira}}, \bibinfo {author} {\bibfnamefont {I.}~\bibnamefont {Jack}},
  \bibinfo {author} {\bibfnamefont {D.~R.~T.}\ \bibnamefont {Jones}}, \ and\
  \bibinfo {author} {\bibfnamefont {C.~G.}\ \bibnamefont {North}},\ }\href
  {\doibase 10.1016/S0550-3213(97)00448-3} {\bibfield  {journal} {\bibinfo
  {journal} {Nucl. Phys.}\ }\textbf {\bibinfo {volume} {B504}},\ \bibinfo
  {pages} {108} (\bibinfo {year} {1997}{\natexlab{a}})},\ \Eprint
  {http://arxiv.org/abs/hep-ph/9705328} {arXiv:hep-ph/9705328 [hep-ph]}
  \BibitemShut {NoStop}%
\bibitem [{\citenamefont {Ferreira}\ \emph
  {et~al.}(1997{\natexlab{b}})\citenamefont {Ferreira}, \citenamefont {Jack},\
  and\ \citenamefont {Jones}}]{Ferreira:1997rc}%
  \BibitemOpen
  \bibfield  {author} {\bibinfo {author} {\bibfnamefont {P.~M.}\ \bibnamefont
  {Ferreira}}, \bibinfo {author} {\bibfnamefont {I.}~\bibnamefont {Jack}}, \
  and\ \bibinfo {author} {\bibfnamefont {D.~R.~T.}\ \bibnamefont {Jones}},\
  }\href {\doibase 10.1016/S0370-2693(97)00291-8} {\bibfield  {journal}
  {\bibinfo  {journal} {Phys. Lett.}\ }\textbf {\bibinfo {volume} {B399}},\
  \bibinfo {pages} {258} (\bibinfo {year} {1997}{\natexlab{b}})},\ \Eprint
  {http://arxiv.org/abs/hep-ph/9702304} {arXiv:hep-ph/9702304 [hep-ph]}
  \BibitemShut {NoStop}%
\bibitem [{\citenamefont {Jack}\ \emph {et~al.}(1997)\citenamefont {Jack},
  \citenamefont {Jones},\ and\ \citenamefont {North}}]{Jack:1996cn}%
  \BibitemOpen
  \bibfield  {author} {\bibinfo {author} {\bibfnamefont {I.}~\bibnamefont
  {Jack}}, \bibinfo {author} {\bibfnamefont {D.~R.~T.}\ \bibnamefont {Jones}},
  \ and\ \bibinfo {author} {\bibfnamefont {C.~G.}\ \bibnamefont {North}},\
  }\href {\doibase 10.1016/S0550-3213(96)00637-2} {\bibfield  {journal}
  {\bibinfo  {journal} {Nucl. Phys.}\ }\textbf {\bibinfo {volume} {B486}},\
  \bibinfo {pages} {479} (\bibinfo {year} {1997})},\ \Eprint
  {http://arxiv.org/abs/hep-ph/9609325} {arXiv:hep-ph/9609325 [hep-ph]}
  \BibitemShut {NoStop}%
\bibitem [{\citenamefont {Kataev}\ and\ \citenamefont
  {Stepanyantz}(2014)}]{Kataev:2013csa}%
  \BibitemOpen
  \bibfield  {author} {\bibinfo {author} {\bibfnamefont {A.~L.}\ \bibnamefont
  {Kataev}}\ and\ \bibinfo {author} {\bibfnamefont {K.~V.}\ \bibnamefont
  {Stepanyantz}},\ }\href {\doibase 10.1016/j.physletb.2014.01.053} {\bibfield
  {journal} {\bibinfo  {journal} {Phys. Lett.}\ }\textbf {\bibinfo {volume}
  {B730}},\ \bibinfo {pages} {184} (\bibinfo {year} {2014})},\ \Eprint
  {http://arxiv.org/abs/1311.0589} {arXiv:1311.0589 [hep-th]} \BibitemShut
  {NoStop}%
\bibitem [{\citenamefont {Shifman}(1996)}]{Shifman:1996iy}%
  \BibitemOpen
  \bibfield  {author} {\bibinfo {author} {\bibfnamefont {M.~A.}\ \bibnamefont
  {Shifman}},\ }\bibfield  {booktitle} {\emph {\bibinfo {booktitle}
  {{Supersymmetry '96: Theoretical perspectives and experimental outlook.
  Proceedings, 4th International Conference, SUSY '96, College Park, USA, May
  29-June 1, 1996}}},\ }\href {\doibase 10.1142/S0217751X96002650} {\bibfield
  {journal} {\bibinfo  {journal} {Int. J. Mod. Phys.}\ }\textbf {\bibinfo
  {volume} {A11}},\ \bibinfo {pages} {5761} (\bibinfo {year} {1996})},\ \Eprint
  {http://arxiv.org/abs/hep-ph/9606281} {arXiv:hep-ph/9606281 [hep-ph]}
  \BibitemShut {NoStop}%
\bibitem [{\citenamefont {Martin}\ and\ \citenamefont
  {Wells}(2001)}]{Martin:2000cr}%
  \BibitemOpen
  \bibfield  {author} {\bibinfo {author} {\bibfnamefont {S.~P.}\ \bibnamefont
  {Martin}}\ and\ \bibinfo {author} {\bibfnamefont {J.~D.}\ \bibnamefont
  {Wells}},\ }\href {\doibase 10.1103/PhysRevD.64.036010} {\bibfield  {journal}
  {\bibinfo  {journal} {Phys. Rev.}\ }\textbf {\bibinfo {volume} {D64}},\
  \bibinfo {pages} {036010} (\bibinfo {year} {2001})},\ \Eprint
  {http://arxiv.org/abs/hep-ph/0011382} {arXiv:hep-ph/0011382 [hep-ph]}
  \BibitemShut {NoStop}%
\bibitem [{\citenamefont {Bajc}\ \emph {et~al.}(2018)\citenamefont {Bajc},
  \citenamefont {Dondi},\ and\ \citenamefont {Sannino}}]{Bajc:2017xwx}%
  \BibitemOpen
  \bibfield  {author} {\bibinfo {author} {\bibfnamefont {B.}~\bibnamefont
  {Bajc}}, \bibinfo {author} {\bibfnamefont {N.~A.}\ \bibnamefont {Dondi}}, \
  and\ \bibinfo {author} {\bibfnamefont {F.}~\bibnamefont {Sannino}},\ }\href
  {\doibase 10.1007/JHEP03(2018)005} {\bibfield  {journal} {\bibinfo  {journal}
  {JHEP}\ }\textbf {\bibinfo {volume} {03}},\ \bibinfo {pages} {005} (\bibinfo
  {year} {2018})},\ \Eprint {http://arxiv.org/abs/1709.07436} {arXiv:1709.07436
  [hep-th]} \BibitemShut {NoStop}%
\bibitem [{\citenamefont {Shtabovenko}\ \emph {et~al.}(2016)\citenamefont
  {Shtabovenko}, \citenamefont {Mertig},\ and\ \citenamefont
  {Orellana}}]{Shtabovenko:2016sxi}%
  \BibitemOpen
  \bibfield  {author} {\bibinfo {author} {\bibfnamefont {V.}~\bibnamefont
  {Shtabovenko}}, \bibinfo {author} {\bibfnamefont {R.}~\bibnamefont {Mertig}},
  \ and\ \bibinfo {author} {\bibfnamefont {F.}~\bibnamefont {Orellana}},\
  }\href {\doibase 10.1016/j.cpc.2016.06.008} {\bibfield  {journal} {\bibinfo
  {journal} {Comput. Phys. Commun.}\ }\textbf {\bibinfo {volume} {207}},\
  \bibinfo {pages} {432} (\bibinfo {year} {2016})},\ \Eprint
  {http://arxiv.org/abs/1601.01167} {arXiv:1601.01167 [hep-ph]} \BibitemShut
  {NoStop}%
\bibitem [{\citenamefont {Vermaseren}(2000)}]{Vermaseren:2000nd}%
  \BibitemOpen
  \bibfield  {author} {\bibinfo {author} {\bibfnamefont {J.~A.~M.}\
  \bibnamefont {Vermaseren}},\ }\href@noop {} {\  (\bibinfo {year} {2000})},\
  \Eprint {http://arxiv.org/abs/math-ph/0010025} {arXiv:math-ph/0010025
  [math-ph]} \BibitemShut {NoStop}%
\bibitem [{\citenamefont {Kuipers}\ \emph {et~al.}(2013)\citenamefont
  {Kuipers}, \citenamefont {Ueda}, \citenamefont {Vermaseren},\ and\
  \citenamefont {Vollinga}}]{Kuipers:2012rf}%
  \BibitemOpen
  \bibfield  {author} {\bibinfo {author} {\bibfnamefont {J.}~\bibnamefont
  {Kuipers}}, \bibinfo {author} {\bibfnamefont {T.}~\bibnamefont {Ueda}},
  \bibinfo {author} {\bibfnamefont {J.~A.~M.}\ \bibnamefont {Vermaseren}}, \
  and\ \bibinfo {author} {\bibfnamefont {J.}~\bibnamefont {Vollinga}},\ }\href
  {\doibase 10.1016/j.cpc.2012.12.028} {\bibfield  {journal} {\bibinfo
  {journal} {Comput. Phys. Commun.}\ }\textbf {\bibinfo {volume} {184}},\
  \bibinfo {pages} {1453} (\bibinfo {year} {2013})},\ \Eprint
  {http://arxiv.org/abs/1203.6543} {arXiv:1203.6543 [cs.SC]} \BibitemShut
  {NoStop}%
\bibitem [{\citenamefont {Cyrol}\ \emph {et~al.}(2017)\citenamefont {Cyrol},
  \citenamefont {Mitter},\ and\ \citenamefont {Strodthoff}}]{Cyrol:2016zqb}%
  \BibitemOpen
  \bibfield  {author} {\bibinfo {author} {\bibfnamefont {A.~K.}\ \bibnamefont
  {Cyrol}}, \bibinfo {author} {\bibfnamefont {M.}~\bibnamefont {Mitter}}, \
  and\ \bibinfo {author} {\bibfnamefont {N.}~\bibnamefont {Strodthoff}},\
  }\href {\doibase 10.1016/j.cpc.2017.05.024} {\bibfield  {journal} {\bibinfo
  {journal} {Comput. Phys. Commun.}\ }\textbf {\bibinfo {volume} {219}},\
  \bibinfo {pages} {346} (\bibinfo {year} {2017})},\ \Eprint
  {http://arxiv.org/abs/1610.09331} {arXiv:1610.09331 [hep-ph]} \BibitemShut
  {NoStop}%
\bibitem [{\citenamefont {Broadhurst}\ \emph {et~al.}(1997)\citenamefont
  {Broadhurst}, \citenamefont {Gracey},\ and\ \citenamefont
  {Kreimer}}]{Broadhurst:1996ur}%
  \BibitemOpen
  \bibfield  {author} {\bibinfo {author} {\bibfnamefont {D.~J.}\ \bibnamefont
  {Broadhurst}}, \bibinfo {author} {\bibfnamefont {J.~A.}\ \bibnamefont
  {Gracey}}, \ and\ \bibinfo {author} {\bibfnamefont {D.}~\bibnamefont
  {Kreimer}},\ }\href {\doibase 10.1007/s002880050500} {\bibfield  {journal}
  {\bibinfo  {journal} {Z. Phys.}\ }\textbf {\bibinfo {volume} {C75}},\
  \bibinfo {pages} {559} (\bibinfo {year} {1997})},\ \Eprint
  {http://arxiv.org/abs/hep-th/9607174} {arXiv:hep-th/9607174 [hep-th]}
  \BibitemShut {NoStop}%
\bibitem [{\citenamefont {Bierenbaum}\ and\ \citenamefont
  {Weinzierl}(2003)}]{Bierenbaum:2003ud}%
  \BibitemOpen
  \bibfield  {author} {\bibinfo {author} {\bibfnamefont {I.}~\bibnamefont
  {Bierenbaum}}\ and\ \bibinfo {author} {\bibfnamefont {S.}~\bibnamefont
  {Weinzierl}},\ }\href {\doibase 10.1140/epjc/s2003-01389-7} {\bibfield
  {journal} {\bibinfo  {journal} {Eur. Phys. J.}\ }\textbf {\bibinfo {volume}
  {C32}},\ \bibinfo {pages} {67} (\bibinfo {year} {2003})},\ \Eprint
  {http://arxiv.org/abs/hep-ph/0308311} {arXiv:hep-ph/0308311 [hep-ph]}
  \BibitemShut {NoStop}%
\bibitem [{\citenamefont {Grozin}(2004)}]{Grozin:2003ak}%
  \BibitemOpen
  \bibfield  {author} {\bibinfo {author} {\bibfnamefont {A.~G.}\ \bibnamefont
  {Grozin}},\ }\bibfield  {booktitle} {\emph {\bibinfo {booktitle}
  {{International Research Workshop on Calculations for Modern and Future
  Colliders (CALC 2003) Dubna, Russia, June 13-21, 2003}}},\ }\href {\doibase
  10.1142/S0217751X04016775} {\bibfield  {journal} {\bibinfo  {journal} {Int.
  J. Mod. Phys.}\ }\textbf {\bibinfo {volume} {A19}},\ \bibinfo {pages} {473}
  (\bibinfo {year} {2004})},\ \Eprint {http://arxiv.org/abs/hep-ph/0307297}
  {arXiv:hep-ph/0307297 [hep-ph]} \BibitemShut {NoStop}%
\bibitem [{\citenamefont {Grozin}(2012)}]{Grozin:2012xi}%
  \BibitemOpen
  \bibfield  {author} {\bibinfo {author} {\bibfnamefont {A.~G.}\ \bibnamefont
  {Grozin}},\ }\href {\doibase 10.1142/S0217751X12300189} {\bibfield  {journal}
  {\bibinfo  {journal} {Int. J. Mod. Phys.}\ }\textbf {\bibinfo {volume}
  {A27}},\ \bibinfo {pages} {1230018} (\bibinfo {year} {2012})},\ \Eprint
  {http://arxiv.org/abs/1206.2572} {arXiv:1206.2572 [hep-ph]} \BibitemShut
  {NoStop}%
\bibitem [{\citenamefont {Huang}\ and\ \citenamefont
  {Liu}(2013)}]{Huang:2012qz}%
  \BibitemOpen
  \bibfield  {author} {\bibinfo {author} {\bibfnamefont {Z.-W.}\ \bibnamefont
  {Huang}}\ and\ \bibinfo {author} {\bibfnamefont {J.}~\bibnamefont {Liu}},\
  }\href {\doibase 10.1016/j.cpc.2013.03.016} {\bibfield  {journal} {\bibinfo
  {journal} {Comput. Phys. Commun.}\ }\textbf {\bibinfo {volume} {184}},\
  \bibinfo {pages} {1973} (\bibinfo {year} {2013})},\ \Eprint
  {http://arxiv.org/abs/1209.3971} {arXiv:1209.3971 [physics.comp-ph]}
  \BibitemShut {NoStop}%
\end{thebibliography}%
\end{document}